\documentclass[12pt,preprint]{aastex}
\usepackage{emulateapj5}
\usepackage{apjfonts}

\begin{document}

\submitted{To appear November 20,2002}

\title{Radio Point Sources and the Thermal SZ Power Spectrum}

\author{Gilbert P. Holder}
\affil{School of Natural Sciences, Institute for Advanced Study,
Princeton, NJ 08540}
\email{holder@ias.edu}

\begin{abstract}
Radio point sources are strongly correlated with clusters of galaxies,
so a significant fraction of the thermal Sunyaev-Zel'dovich (SZ) effect
signal could be affected by point source contamination. Based on
empirical estimates of the radio galaxy population, 
it is shown that the rms temperature 
fluctuations of the thermal SZ effect could be underestimated by as much 
as 30\% at an observing frequency of $30$ GHz at $\ell \ga 1000$. 
The effect is larger at higher multipoles. 
If the recent report of excess power at small angular scales is to be
explained by the thermal SZ effect, then radio point sources 
at an observing frequency of 30 GHz must be a surprisingly weak
contaminant of the SZ effect for low-mass clusters.
\end{abstract}

\section{Introduction}

Recent and upcoming measurements of anisotropies in the cosmic microwave
background (CMB) may be sensitive enough (a few $\mu K$) and at high
enough angular resolution (a few arcminutes) to detect fluctuations in the
microwave background due to unresolved distant and faint clusters of
galaxies, due to the thermal Sunyaev-Zel'dovich (SZ) effect.  
As the largest expected signal of ``secondary'' anisotropies
(anisotropy arising from processes at low redshift),
such a detection will mark an important milestone, as well as provide
important constraints on cosmological parameters and the thermal history
of intra-cluster gas.

Significant work has gone into the problem of extracting the signal of
the ``primary'' anisotropies (those imprinted at $z \sim 1100$) from the
expected foreground sources of contamination, such as dust, star-forming
galaxies, and radio point sources. A major aid in this extraction is the
lack of correlation between the signal and the contaminant. In searching
for the signal of secondary anisotropies, this is no longer the case. While
this does not provide any deep or profound problems, it does provide
reason for caution in interpretation of detected signals in the presence
of foregrounds.  

It is well-known
that clusters of galaxies observed at frequencies near
30 GHz (wavelength 1 cm) are likely to contain radio point
sources \citep{birkinshaw99}, 
primarily due to radio emission from the cluster galaxies 
themselves \citep{cooray98}.
Therefore, the signal of the thermal SZ effect can be significantly diluted,
depending on how these point sources are handled. There are two
possible modes of dilution: removal of point sources
can lead to removal of SZ signal, and point sources that are
not removed can fill in an SZ decrement.  In this work
we estimate the expected dilution of the thermal SZ effect signal at
an observing frequency of 30 GHz, appropriate for an experiment such 
as CBI \citep{padin01}.  

While the focus in this paper will be contamination of the thermal
SZ effect by radio point sources, at higher frequencies dusty starbursting
galaxies could be important contaminants of the SZ signal \citep{blain98},
due to gravitational lensing effects. These same contaminants will be
troublesome for signals of lensing in the CMB. Multifrequency observations
alleviate these concerns somewhat, but the uncertain and heterogeneous 
spectral behavior of both radio and submm point sources present a challenge
for precise measurements of secondary anisotropies.

The thermal SZ effect
and its angular power spectrum is reviewed in \S\ref{sec:sz}. 
In \S\ref{sec:radio} we outline the relevant statistics of 
radio point sources in galaxy clusters and apply them to
the thermal SZ power spectrum in \S\ref{sec:effects}. 
The implications of these results are discussed in \S\ref{sec:results}.

\section{Thermal SZ From Galaxy Clusters}
\label{sec:sz}

The thermal SZ effect \citep{sunyaev72,birkinshaw99,carlstrom00} arises
from Compton scattering of cool CMB photons with hot electrons in the
deep potential wells of galaxy clusters. 
A striking feature of the
SZ effect is its unique spectrum. Relative to the CMB, the SZ effect
manifests itself as a deficit of photons at frequencies below about
218 GHz and as an excess at higher frequencies. The decrement at
low frequencies is particularly useful, since there are very few
astronomical signals which show up as ``holes'' in the sky.

The temperature decrement (or increment), ignoring relativistic corrections
\citep{rephaeli95}, is given by
\begin{equation}
 \frac{\Delta T_{SZ}}{T_{CMB}} =  f(x) \ y  = f(x) \int
n_e \frac{k_B T_e}{m_e c^2} \sigma_T \, d\ell,
\label{eqn:dT}
\end{equation}
where $y$ is the Compton $y$ parameter, $n_e$ is the electron number density,
$T_e$ is the electron temperature, and 
$x \equiv \frac{h\nu}{k_B T_{CMB}}$ is the observing frequency in natural
units.  The frequency dependent factor is given by
$f(x) = \left(x \frac{e^x+1}{e^x-1} -4\right)$. When integrated over the
entire angular extent of the cluster, the SZ flux has a scaling of
$S_{sz} \propto d_A^{-2} f_g h\,M\,T_e$, with $d_A$ the angular diameter 
distance in units of $h^{-1}$Mpc, 
$f_g$ the gas mass fraction, $h$ the Hubble constant in units
of 100 km\,s$^{-1}$Mpc$^{-1}$ and $M$ is the cluster mass in units 
of $h^{-1}M_\odot$. 
Assuming the virial relation $T\propto M^{2/3}$ gives the SZ flux scaling
with mass as $M^{5/3}$.

The primary anisotropies in the CMB are thought to be Gaussian in nature,
and can therefore be characterized entirely through the angular power
spectrum. This is not true for either foregrounds or secondary anisotropies.
The angular power spectrum is therefore 
not nearly as useful, but it does present
a common language with which to work and we will adopt it here. 

Adopting the flat-sky approximation, the usual multipole expansion becomes a 
two dimensional Fourier transform, with $\ell = 2 \pi R_{uv}$, where 
$R_{uv} \equiv \sqrt{u^2+v^2}$
is the radial distance in the Fourier plane and $u$ and $v$ are the 
Fourier conjugate variables to $\theta_x$ (e.g., R.A.) 
and $\theta_y$ (e.g., Dec.) on the sky. 
The variance of the Fourier amplitudes at radius $R_{uv}$ is equal to
$c_\ell$. Denoting the Fourier transform of the cluster SZ 
temperature decrement profile as
$\tilde{T}$, the angular power spectrum of a galaxy cluster at the
center of a field is simply
$c_\ell = \tilde{T}^2$, assuming azimuthal symmetry for the cluster.

Spatial correlations between galaxy clusters are negligible for $\ell \ga 100$
\citep{komatsu99a} and in this work we are interested in $\ell \ga 1000$, so
the correlations can be safely neglected. 
The angular power spectrum can therefore be thought of as Poisson shot noise, 
where each ``shot'' has an angular profile.  For a Poisson process, at each
position in the Fourier plane a randomly placed source will have an amplitude
$\tilde{T}$ but will have a random phase. The collection of sources will
thus constitute a random walk of the Fourier amplitude with step size
$\tilde{T}$, leading to a final $c_\ell = N \tilde{T}^2$, where N is the
number of galaxy clusters.

As an integral over redshift and cluster mass, this can be expressed as
\begin{equation}
c_\ell (\ell)  = \int_0^\infty dz {d_A(z)^2 (1+z)^2 \over H(z)}
\int_0^\infty d \ln{M} \, \tilde{T}(M,z,\ell)^2 \, {dn \over d \ln{M} }, 
\label{eqn:cl}
\end{equation}
where $d_A(z)$ is the angular diameter distance, $H(z)$ is the
Hubble constant, and $dn/d\ln{M}$ is the differential comoving number density 
per log interval in mass. A very fast method to
obtain $\tilde{T}$ for somewhat realistic cluster profiles with azimuthal 
symmetry is to take advantage of fast Hankel transform routines that 
exist \citep{anderson82}. Note that here, and everywhere below, we work in
units where $c_\ell$ has units of $\mu K^2$. 

The power spectrum calculation exactly follows the procedure
of Holder and Carlstrom (2001) \nocite{holder01}. 
We choose cosmological parameters
$\Omega_m=0.3$, $\Omega_\Lambda=0.7$, $\sigma _8=1$, $h=0.7$, 
$\Omega_{b}h^2 =0.02$, $n=1$, and zero neutrino mass.
The differential comoving number density is adopted from recent fits
to large numerical simulations of structure formation \citep{jenkins01},
and is a function of the variance on mass scale $M$. This variance 
was calculated using the power spectrum for our adopted cosmological
model derived from the fitting functions of Eisenstein and Hu (1999)
\nocite{eisenstein99a}.

As a cluster model, we adopt the simple toy model of Holder and Carlstrom
(2001), with a density profile of the form 
$n_e \propto 1/(r_c^2+R_v^2)$, with $r_c$ a core radius
and $R_v$ the virial radius, derived from the spherical collapse model
\citep{lahav91}.  The relation between core radius and virial radius is taken
to be a constant value (ten), roughly as would be expected
for the case of self-similar evolution of the cluster population.
The gas temperature as a function of mass
was taken from hydrodynamical simulations \citep{bryan98} and the
gas was assumed to be isothermal. 

For this work, we are interested in the {\em relative} effects of radio 
point sources on the SZ angular power spectrum, so the details of the
power spectrum are not crucial.  We have verified that the results below are
robust to the choice of model for generating the SZ power spectrum.
The angular power spectrum resulting from
our recipe is in broad agreement with results from large cosmological
hydrodynamical simulations \citep{springel00}. Temperature
gradients will affect the details of the shape of the peak of the power
spectrum, while shifting the normalization of the mass temperature relation
introduces a direct scaling of the amplitude of the angular power spectrum.
For example, using the observed normalization of the mass-temperature relation
\citep{finoguenov01} leads to an rms temperature fluctuation that is 
approximately 50\% higher. 
Given the uncertain relation between the observed X-ray 
temperatures and the (most relevant for our purposes) 
mean electron temperature, we choose to use the normalization from simulations.

Putting the pieces together, the integrand of Equation \ref{eqn:cl} is
shown as a function of mass and redshift in Figure \ref{fig:dcl}.
Self-similar evolution of the intra-cluster medium
was assumed, but the gas evolution 
history has very little effect on such a plot. At $\ell=1000$, most of
the signal is coming from somewhat massive, relatively nearby clusters,
while the dominant contribution to higher $\ell$ is coming from distant
low-mass clusters, with a significant tail extending to $z=2$. 

It is easy to see why radio point sources could be a problem. 
At $\ell=1000$, the signal is coming from nearby clusters, where each point
source should be relatively bright, and from fairly massive clusters, which
have more cluster members and therefore could be expected to have 
more point sources. These clusters also have stronger SZ emission
and are more extended on the sky, which will somewhat mitigate
the point source contamination. At higher $\ell$, the clusters have fewer point
sources and the ones that they do have are diluted by the luminosity
distance. However, the cluster signal is weaker, making point sources
relatively more important. Thus, over the whole range of the peak of
the SZ angular spectrum it is expected that radio point source
contamination might be important for experiments at low frequencies.


\centerline{{\vbox{\epsfxsize=8cm\epsfbox{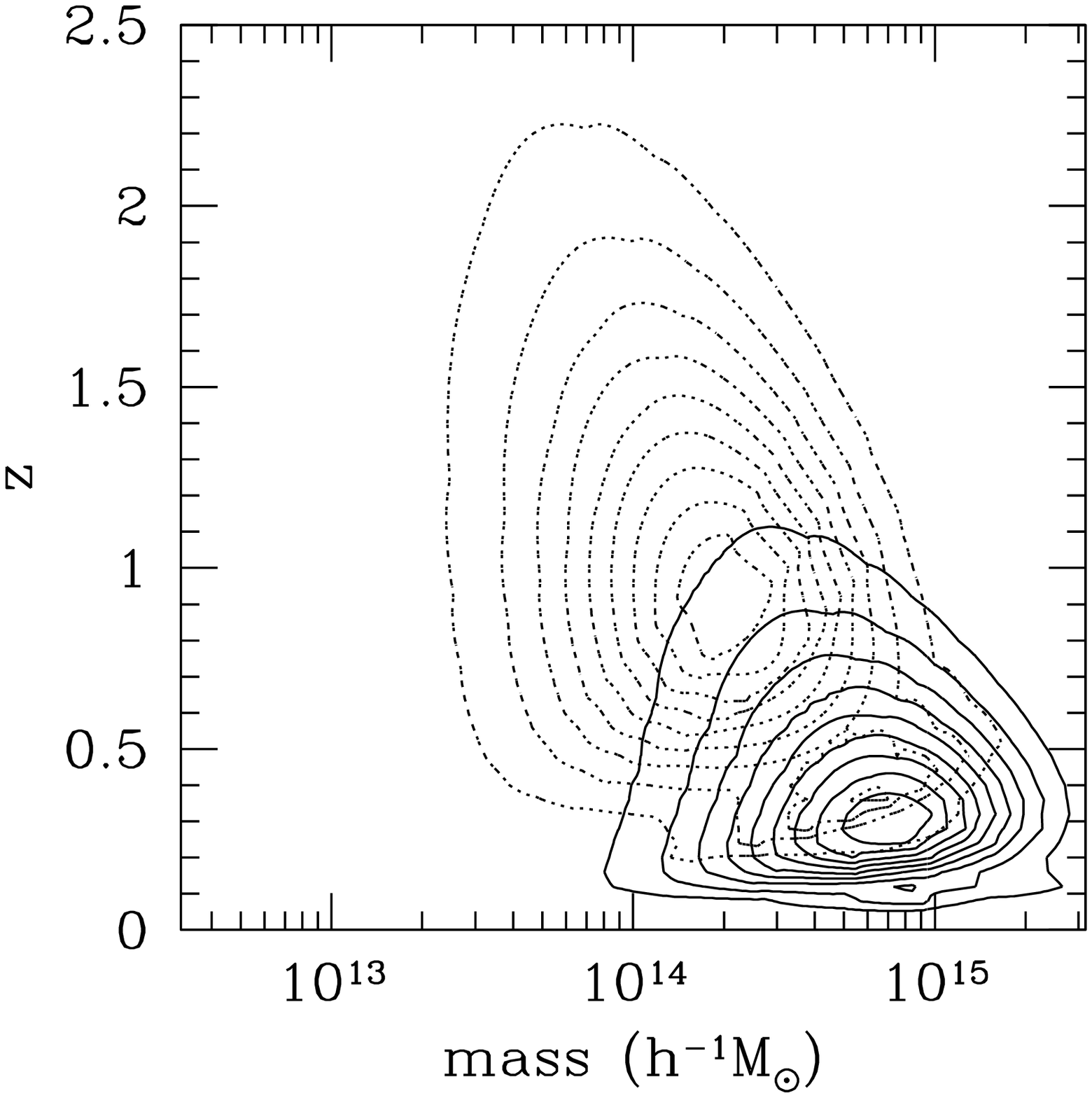}}}}
\figcaption{Relative differential contribution to the angular power spectrum, 
$dc_\ell / d \ln{M}dz$, normalized to peak contribution. Solid contours
show contributions to $\ell=1000$, with contour levels of 10\% of the
peak contribution. Dotted contours show contributions to $\ell=4000$, also
with contour levels of 10\% of peak.
\label{fig:dcl}}

\section{Radio Point Sources in Galaxy Clusters}
\label{sec:radio}

Radio point sources are well measured and catalogued in terms of their
flux distribution and source density at an observing frequency of 1.4 GHz,
due to the NVSS \citep{condon98} and FIRST \citep{white97} surveys. 
Unfortunately, radio point sources often have non-trivial spectra
\citep{herbig92}, so it is not easy to simply extrapolate from one
frequency to another. The mean spectral index from 21 cm to 1 cm is
approximately $\alpha =-0.7$, where $S(Jy) \propto \nu^\alpha$
\citep{cooray98} but there is significant dispersion in the observed 
spectral indices \citep{cooray98,taylor01}. 

At 1 cm, there are no large scale deep surveys for point
sources that can be used to accurately characterize the point source
population, but there are a large number of pointed observations toward
clusters for SZ effect observations \citep{carlstrom99} at this
wavelength. The majority of observed clusters has at least one
point source with a flux at 1 cm greater than 1 mJy.  When compared
with the point source abundance in fields not containing clusters, it
is clear that most point sources in fields with galaxy clusters
must be physically associated with the galaxy
clusters \citep{cooray98}, most likely the galaxy cluster members.

Radio emission from galaxy cluster members has been studied in detail
\citep{ledlow96} at 21 cm ($\nu=1.4$ GHz), with the fraction of
galaxies at a given radio luminosity (per log interval) 
nearly flat at radio powers (at 21 cm)
below $10^{24.8}$ W\,Hz$^{-1}$ and falling quickly above this power.  
To the optical and radio flux limits of their survey, 
roughly 10\% of cluster members showed
some amount of point source emission. Assuming a typical spectral index
$\alpha=-0.7$ and our fiducial cosmology, the break power at
21 cm corresponds to an observed flux at 1 cm of 7 mJy for a source
at $z=0.2$ and 1 mJy for a source at $z=0.5$. This flux level is 
in rough agreement with the typical point source fluxes observed by the
OVRO/BIMA SZ imaging experiment \citep{carlstrom99, reese02}. If no point
source subtraction of any kind were done, these point sources would
be a non-negligible fraction of the thermal SZ signal from massive clusters
($\sim 10^{15} h^{-1}M_\odot$). 

The nearly flat probability distribution (in log flux)
with point source flux leads to
the typical galaxy cluster having the brightest point source dominating the
total radio flux.  
From an observational standpoint this is desirable, since
an SZ experiment only needs to remove a bright point source or two to be
confident that the measurement is not contaminated by radio emission from 
galaxy cluster members. 

The expected integrated thermal SZ signal for a 
$10^{15} h^{-1}M_\odot$ at $z \sim 0.5$
is on the order of 10-20 mJy (at 30 GHz) \citep{holder01}, and should scale
as $M^{5/3}$. This is in agreement with current observations 
\citep{carlstrom99},
 but current data does not extend over a large range in mass. 
The radio point source flux should scale approximately with
the number of galaxies, which scales as the mass \citep{carlberg96}. A
typical point source flux for massive galaxy clusters at this redshift 
is on the order of 1 mJy at 30 GHz and is 
therefore roughly 5\% of the total thermal SZ flux. The 
relative importance of point sources should therefore scale with mass as
$M^{-2/3}$, while the redshift evolution of the relative importance will scale
as $(1+z)^{-3}$ for a flat spectral index. None of these numbers are
particularly well constrained observationally, but these estimates constitute 
our baseline model. 

\section{Effects of Point Sources on the SZ Power Spectrum}
\label{sec:effects}

At some level, all clusters have radio point sources. Ideally, one
would be able to identify and remove all point sources in the field
of view with a beam size that is matched to the size of the point source
and much smaller than the extent of the SZ signal.
In such a case, the amount of SZ flux removed would be negligible. 
Not surprisingly, this is exactly the attempted strategy of choice for 
interferometric SZ experiments, such as
the Ryle Telescope \citep{jones93} and OVRO/BIMA \citep{carlstrom96}, where
high angular resolution measurements are performed simultaneously for
point source removal. 
It is not feasible to detect all point sources in a field,
but such methods can easily remove point sources to a flux level below
$10\%$ of the peak SZ flux, with an effective resolution of 
roughly 10'' or better. With such a strategy, the residual effect of point 
sources on the SZ power spectrum would be negligible. 

Such a strategy is not currently feasible for CMB experiments such 
as CBI. As primarily a CMB experiment, CBI
is not concerned with point sources that do not contribute more than
about 10 $\mu K$ to the rms temperature fluctuations. This translates
into a flux threshold of a few mJy at 30 GHz. Clusters with masses
near $10^{14} h^{-1} M_\odot$ have a total SZ flux of a mJy or less
\citep{holder01}, so a 1 mJy point source could be problematic.
If all clusters had point sources near 1 mJy,
effectively no clusters below $\sim 2 \times 10^{14} h^{-1} M_\odot$ 
in figure \ref{fig:dcl} would contribute to the anisotropy, with 
a somewhat reduced contribution from slightly larger clusters. At
$\ell=1000$ the difference would be noticeable but at $\ell=4000$
it might be expected that as much as half of the power could be missing.

There are two leading strategies for dealing with point sources. One
strategy is to identify possible point sources in catalogs at
21 cm and follow up with pointed observations at high angular resolution.
Another strategy is to combine the data in a way that it is insensitive
to any amount of flux coming from positions of known point sources,
known as a constraint matrix approach \citep{bond98}. Both
methods are susceptible to point sources with ``inverted'' spectra, where
the flux is higher at higher frequencies, and therefore could be
missed in the 21 cm catalog. Such inverted sources are rare, and are not
expected to be a dominant source of error. For extraction of primary 
CMB anisotropies both methods work quite well \citep{padin01,halverson01},
but for the SZ effect, point source subtraction will remove all SZ flux 
within the point source subtraction beam area from the map at the position 
of the point source. 

\centerline{{\vbox{\epsfxsize=8cm\epsfbox{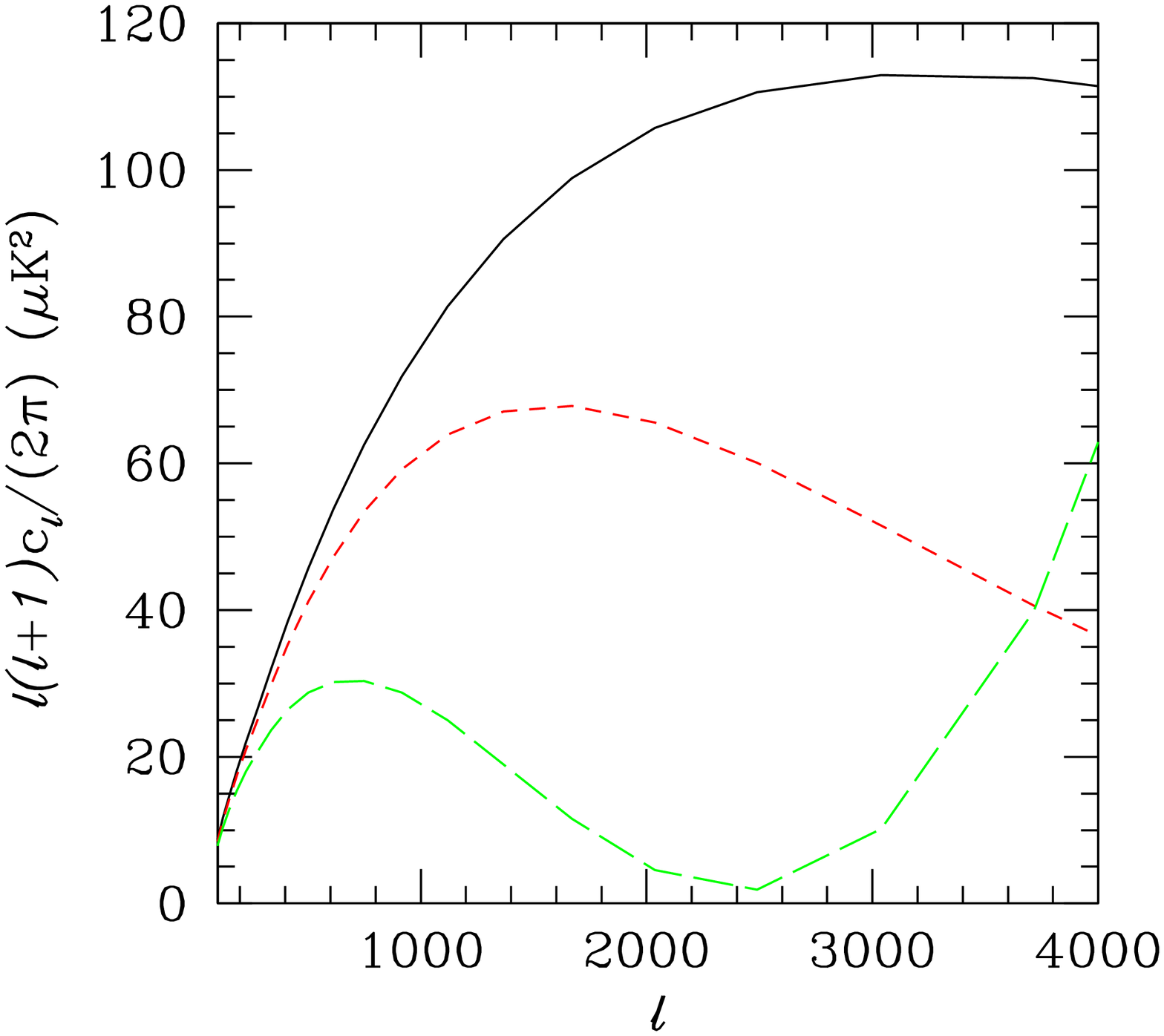}}}}
\figcaption{Effects of point source subtraction on SZ angular power spectrum. 
Solid line shows true power 
spectrum, short dashed line indicates power spectrum after central
1' has been removed from all clusters. 
Long dashed lines indicate resulting power spectrum after removing central 
region of all clusters with a constraint matrix approach. This represents
a worst-case scenario of all galaxy clusters hosting bright point sources.
\label{fig:cl_pts_1}
}

Using the constraint matrix method, the data are combined such that any
constant amplitude signal in the Fourier plane (with the phase center at
the position of the point source) does not contribute to the measured
CMB anisotropy. This will remove from the measured power
the signal from a coincident cluster, averaged over the data. Since the 
CMB measurements are sensitive primarily to scales of several arcminutes
and higher, this will effectively remove much of the cluster signal. 
Approximately, the constraint matrix should project out of the data
the mean cluster signal in the Fourier plane.

As a toy model, we look at the effects of removing a point source from the 
center of {\em all} clusters. In figure \ref{fig:cl_pts_1} we show the effects
of point source subtraction for the case of a central point source that has
either been removed with a 1' (FWHM) beam or had power from the center 
of each cluster projected out of the data by a constraint matrix approach.
In the first case, we multiplied the SZ profiles by the appropriate beam
and subtracted the result from the initial profile.  This is 
correct only for an experiment with a reference beam that is outside the
cluster, which is rarely strictly true, but sufficient for our purposes.
For the second
case, we calculate the mean amplitude of the cluster profile
(in the Fourier domain) weighted by $\ell$ to mimic the effects of a uniform
window function. We assume $\ell$ coverage between $\ell=500-4000$ and
assume uniform coverage of the Fourier plane between these values. 

The constraint matrix approach severely underestimates the SZ power, since
it is effectively doing point source removal with the beam size set by the
approximate angular resolution of the CMB measurements. Since this is
larger than 1', it is to be expected that such an approach will remove
significantly more power than direct subtraction with a 1' beam. The 
anisotropy power from galaxy clusters coincident with point sources
is effectively ``nulled out'' because of the relatively large synthesized
beam of CBI.

\centerline{{\vbox{\epsfxsize=8cm\epsfbox{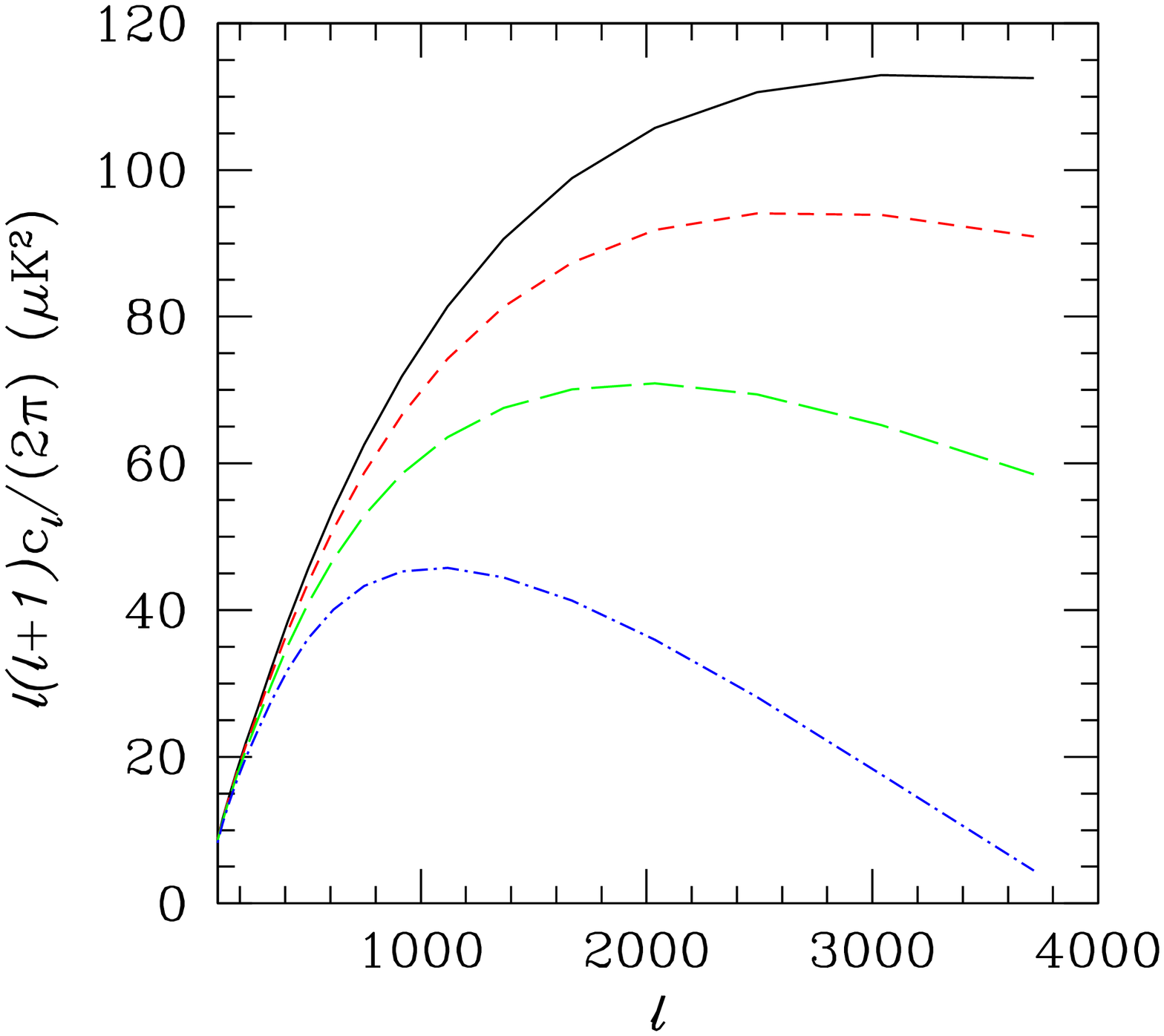}}}}
\figcaption{Effect of unsubtracted correlated 
point sources on SZ angular power spectrum
for three levels of point source contamination. Top to bottom: no
contamination (solid), point source flux 2\% of SZ flux for z=0.5 massive
cluster (short dashed), point source flux 5\% (long dashed), point
source flux 10\% (dot-dashed). This is an approximation to the effects
of no point source removal. Note that the power expected from point
sources (which goes as $\ell^2$) has been subtracted, leaving only
the SZ power spectrum and the effects of the correlation.
\label{fig:cl_pts_2}}
\vskip 0.2in

Direct subtraction can lead to an underestimate of the power 
of roughly 50\% on scales of $\ell=3000$. Therefore, a
detection of a signal of amplitude $100 \mu K^2$ would correspond to a true
SZ angular power spectrum of $200 \mu K^2$. Improving the angular resolution
of the source subtraction helps considerably, as would using a reference
beam within the cluster, as some of the SZ signal would be in both the
source beam and the reference beam and would not be subtracted. For this
reason, near future interferometers such as the SZA and AMI, 
with large spacings for point source removal, should not have
severe problems with point source contamination.

The worst-case scenario is that {\em every} point source that CBI
projected out of the data was at the center of a galaxy cluster and that
every galaxy cluster had a point source at the center.  With nearly 100
NVSS sources per square degree and only about 20 clusters above
$10^{14} h^{-1}M_\odot$ per square degree \citep{holder01} this only 
requires one in five sources to be located in clusters.  A cursory check
of NVSS images of known Abell clusters rules out this hypothesis being
correct, but it is not impossible that a significant fraction of galaxy 
clusters were excluded in this way. If half of the SZ signal is coming from
clusters coincident with known point sources, CBI would be missing about 
half of the fluctuation power from the thermal SZ effect.

In practice, point source subtraction will only be performed to a limiting
flux level, meaning that many clusters will not have source subtraction
done. In this case, the remaining point sources will partly fill in the
SZ decrement. The extreme case is where none of the point sources are
subtracted, leaving all point sources sitting in galaxy clusters, filling
in the SZ decrement and reducing the fluctuation power. In fact, for the
clusters contributing the bulk of the SZ signal (see Figure 1), this
is most likely a good approximation to CBI. 

As a simple model, we assume the scalings with mass from section 
\S\ref{sec:radio}, with some additional constraints. We assume that
all relevant nearby point sources are removed efficiently, and only
include point source contamination for clusters with $z>0.2$. This will
slightly underestimate the effect of radio point sources for two
reasons. There
will invariably be some nearby clusters with relatively faint point
sources that have escaped detection, and at the same time any point source
subtraction, as shown above, can remove a significant amount of SZ flux.
We also assume that only clusters with $M>10^{13}h^{-1}M_\odot$ contain
radio sources.
We assume a canonical typical point
source flux of 5\% of the total SZ flux 
for a $10^{15}h^{-1}M_\odot$ cluster at $z=0.5$.

In the Fourier domain, point source amplitudes are completely
correlated at each point, and for a central point source the signal
will be entirely real. At any point in Fourier space, the combined
signal from the cluster and unresolved point sources will be 
$S_{net} \equiv (\tilde{T} - S_{pt})$.  This will contribute to 
$c_\ell$, on average, 
$S_{net}^2 = c_{\ell;no} + <S_{pt}^2> - 2\tilde{T}<S_{pt}>$, where
$c_{\ell;no}$ is the contribution in the absence of residual unsubtracted
point sources.  If the point sources are not concentrated at the galaxy
cluster center (where they would be if the emission is mainly from the
bright galaxies), but instead trace the gas we would expect $<S_{pt}>$
in the Fourier plane to have the same shape as $\tilde{T}$, rather than
being constant. For simplicity, we assume point sources are strongly
centrally concentrated and place them at the cluster centers.

In figure \ref{fig:cl_pts_2} we show the SZ power spectrum that would be
inferred. The power spectrum due to the
point sources alone has been subtracted. 
Unsubtracted point sources cause
a significant underestimate of the SZ angular power spectrum. 
Assuming a larger contamination of 10\% of the SZ flux for a $z=0.5$
$10^{15}h^{-1}M_\odot$ eliminates nearly all of the SZ power, while
dropping the contamination to only 2\% reduces the missing power to only
about 20\%. 

\section{Discussion}
\label{sec:results}

Point source contamination makes interpretation of detection of fluctuations
in the CMB from the thermal SZ effect very difficult. To compare
these fluctuations to predictions from either semi-analytic modeling or
numerical simulations, some treatment of the effects of point source
subtraction (or non-subtraction) is required. This would
require a recipe for the cluster radio galaxy populations. Current
predictions of the thermal SZ power spectrum are almost certainly
overestimating the thermal SZ power that CBI could observe.

The measured power at 30 GHz at high multipoles could
be less than 30\% of the true SZ power, but more likely
is measuring roughly 50-75\%. 
Specifically, the recent 
tentative report of temperature fluctuations at high multipoles of
close to $500 \mu K^2$ \citep{mason02}, if due to the thermal SZ effect would
indicate a true signal on the sky of roughly $1000 \mu K^2$. 
A temperature {\em rms} of $25 \mu K$ represents a bit of a challenge for 
theoretical models \citep{bond02,komatsu02}, so a true signal greater than 
$30 \mu K$ would require a major rethinking of the physics of galaxy clusters 
and/or some fine tuning of cosmological parameters. 
The simplest way to increase the expected SZ power is
to increase $\sigma_8$. A true SZ signal of more than $30 \mu$K would suggest
$\sigma_8 \ga 1.1$ \citep{komatsu02}, a value that is not preferred by
current CMB data \citep{bond02} but is not ruled out. From figure 3 of
\citet{bond02} such a high value could most easily be accommodated if the
Hubble constant were significantly lower than the value suggested by
HST measurements \citep{freedman01}.

Alternatively, if the measured power at high $\ell$ really is a measurement
of the thermal SZ effect, this would be evidence that low-mass clusters at
$z\sim0.5$ are remarkably devoid of bright radio sources. This suggests
that upcoming SZ surveys at 30 GHz (SZA) or 15 GHz (AMI) should have 
surprisingly clear extragalactic skies.

The statistics of point sources at high radio frequencies are very poorly
constrained, making detailed predictions of point source contamination
difficult. We have adopted an approach to modeling point source contamination
of SZ signal that is empirically motivated, with only the relative
importance of point sources to SZ signal at a single mass scale as a free
parameter.
Upcoming SZ/CMB experiments with high angular resolution,
specifically for the purpose of point source detection, such as AMI and
SZA will provide a wealth of information on radio point sources at these
frequencies, while at the same time providing valuable information on the
SZ effect from galaxy clusters.  

Single frequency measurements of the fluctuations due to the thermal SZ
effect at frequencies below $\sim$90 GHz that
do not subtract point sources with a small beam to a fairly low flux level
will be contaminated at a largely unknown but almost certainly
significant level. Higher frequency measurements could have similar problems
from dusty starburst galaxies, but most of these sources are expected to not
be associated with the galaxy cluster members. While lensing effects lead to
an enhancement  of the confusion noise \citep{blain98}, 
it does not lead, on average to an increased average flux. 

Experiments with multiple frequencies will be required for a robust 
determination of the amplitude of the thermal SZ signal, and the strong
correlations between radio point sources and galaxy clusters and
submm point sources and galaxy clusters (primarily due to gravitational
lensing) will require careful attention. 

The general problem of correlation between secondary anisotropies and 
foregrounds will be increasingly important. The thermal SZ effect
may be the most significant example, but correlations between
the lensing of the CMB, for example, and radio and submm
point sources will reduce the expected signal and/or modify the noise 
properties of any attempted reconstructions.
Clearly, a better understanding of the covariance between 
various foregrounds and backgrounds and secondary anisotropies of interest 
will be required if such signals are to be used as useful tests of our
understanding of cosmology and structure formation.

\acknowledgements{This work was supported by a W.M. Keck Fellowship at the IAS.
I would like to thank Erik Reese for many useful conversations.
}

\end{document}